%% file: DS1597.tex
\documentclass{aa}
\input psfig.sty
\footnotesize
\newdimen\digitwidth    
\setbox0=\hbox{\rm0}
\digitwidth=\wd0
\catcode`!=\active
\def!{\kern\digitwidth}
\normalsize

\usepackage{rotating}

\begin{document}

 \thesaurus{20     
		(08.16.6;  
		 02.16.2;  
		 04.19.1)  
		--- DS1597	
	    }

 \title{Pulsars identified from the NRAO VLA Sky Survey}

 \author{J.L. Han\inst{1,2,3}, W.W. Tian\inst{1,3}
         }
 \offprints{J.L. Han}
 \institute{ Beijing Astronomical Observatory, Chinese Academy of Sciences (CAS),
	     Beijing 100012, China 
 \and 	     Beijing Astronomy Center, CAS-PKU, Beijing 100871, China 
		(jhan@bac.pku.edu.cn)
 \and        Max-Planck-Institut f\"ur Radioastronomie, Auf dem H\"ugel 69,
             D-53121 Bonn, Germany	   }
\date{Received 1998 June 26; accepted 1999 March 3.
 {\bf A\&A Supplement Series}
      }
 \maketitle

\begin{abstract}
We identified 97 strong pulsars from the NRAO VLA Sky Survey (NVSS) at 1.4
GHz north of Dec(J2000) $>-40\degr$. The total flux density,
linear polarization intensity and polarization angle (PA) of all pulsars
are extracted from the NVSS catalog. The well-calibrated PA measurement of
5 pulsars can be used for absolute PA calibrations in other observations.
Comparing the source positions with those in the pulsar catalog,
we got the first measurement of the proper motion upper limit of PSR
B0031$-$07, which is $\mu_{\alpha}\cos\delta = -102\pm74$ mas yr$^{-1}$
and  $\mu_{\delta} = -105\pm78$ mas yr$^{-1}$.

\end{abstract}

\section{Introduction}
Compared with other types of radio sources, pulsars are known to have
strong polarization, even up to 100\% if one observes them with high
time resolutions. Pulsar polarization would
be smeared somehow if they are observed as continuum point sources over a
duration much longer than a pulsar period, mainly because of the
fast swing of polarization angle across a pulse profile. However, we will
show in this paper that is not so serious as generally believed.

Pulsars have high (birth) velocities, on average $~$450 km~s$^{-1}$ (Lyne \&
Lorimer  \cite{ll94}) and maybe up to 1600 km~s$^{-1}$ for individuals 
(e.g. Cordes \& Chernoff \cite{cc98}), much faster than that of other types
of stars (typically a few tens km~s$^{-1}$). The high velocity was probably
caused by the asymmetric kick during supernova explosion when a pulsar
was born. This leads to a large proper motion for (nearby) pulsars. 
However, measuring the proper motion is not an easy task since the
precise positions of a pulsar at well-seperated epochs have to be
precisely measured. Up to now, there are 96 pulsars with proper motion
measurements (e.g. Taylor, Manchester \& Lyne \cite{tml93}; Fomalont et al.
\cite{fomet97}).

Recently, the National Radio Astronomical Observatory (NRAO) Very Large
Array (VLA) Sky Survey (NVSS) has been finished, which covers the sky
north of Dec(J2000) $= -40\degr$ at 1.4 GHz (Condon et al. \cite{conet98}).
The survey detected more than 1.8 million sources, with polarization
measurements, down to a flux density limit about 2.5 mJy. Observations
have a resolution of $45''$, but the positional accuracy is a few arcsec
for weak sources, and much better for strong sources. The observations
were made with two IF channels at 1.365 and 1.435 GHz with an effective
bandwidth of 42MHz each. Most sources in the NVSS were observed in three
pointings of 23 sec each. The final sky map is the weighted sum from these
pointings (Condon et al. \cite{conet98}). 

We had tried to identify the pulsars from the NVSS catalog, and then to
investigate the pulsar polarization properties and proper motions from
continuum observations. In the sky region covered by NVSS,
there are 520 known pulsars according to the updated pulsar catalog of 
Taylor, Manchester \& Lyne (\cite{tml93}. Updated catalog was kindly
provided by Manchester). Using the latest version of the NVSS catalog
(with 1814748 entries), we identified 97 strong pulsars according to
positional coincidence. 
During revising this paper for publication, we noticed that similar
identification work has been done by Kaplan et al. (\cite{kapet98}), but
they emphasized the other aspects, such as position accuracy, scintillation
effects and completeness of detections. Comparing to Kaplan et al. 
(\cite{kapet98}), we got 24 further new identifications. In the following,
we will not repeat their work, but present our results in Sect.2. We discuss
briefly  in Sect.~3 about scintillations (Sect.~3.1),  pulsar
polarization properties (Sect.~3.2), and proper motions (Sect.~3.3).
We compared the pulsar positions with those from the pulsar catalog if
the epochs were
seperated over more than 5 years, and got the upper limits of proper motion
of 18 pulsars, including one pulsar which has had no proper motion measurements
previously.

\section{Identification and Results}

We took positions of pulsars from the updated catalog of Taylor et al. 
(\cite{tml93}). PSR names in J2000, and B1950 if applicable, are given
in the Columns (1) and (2) of the Table 1. Their positions are given in
columns (3) and (4), generally with an accuracy better than $0.1''$, but
occasionally up to a few arcsec. These positions were determined by timing
observations or interfermetric measurements at epoch for the
position\footnote{
There are two epochs in the pulsar catalog, one (``pepoch'') for pulsar period
and period derivatives and the other (``epoch'') for pulsar position. If
``epoch'' was not available, we used the ``pepoch'' as instructed
by Manchester (private communications).}
in Column (5).
For comparison, we list in column (6) the flux density at 1.4 GHz from
pulsar catalog, which were normally obtained from the average of several
pulsar observation sessions to overcome scintillation effects. We searched
for radio sources in the NVSS catalog within $30''$ angular distance 
around each of the 520 pulsar positions. Only 106 radio sources were
found to match the positions and are probably related to pulsars. The
positions of  the NVSS sources are listed in columns (7) and (8).
The angular offset from pulsar positions ``$\Delta$'' in arcsec
is given in column (9). The flux density and polarization parameters of
the NVSS sources extracted from the NVSS catalog are listed in columns
(10)--(13). A blank in these columns indicates no significant detection
above the sensitivity limit of linear polarization of the NVSS ($\sim$0.5
mJy). We marked in column (14) if there was any further consideration
during identifications.

Note that the epochs for pulsar position in the pulsar catalog differ
from that of NVSS observations. However, even if a pulsar has the largest
proper motion, eg. 400 mas per year, then after 20 years, the position
offset would be $8''$. So, our search in $30''$ should not miss any known
pulsar if it is detectable by the NVSS.\footnote{
We missed 4 pulsars which appear in Table 1 of Kaplan et al. 
(\cite{kapet98}): PSRs B1823-11, B1900-06, B1901+10, and B2323+63. Their
position offsets to the NVSS sources or position uncertainties are too
large ($>30''$) to make significant assessment. For the same reason,
we removed J1848+0651 from our sample which was included by
Kaplan et al. (\cite{kapet98}).}

On the other hand, the NVSS was done over a long period, from $\sim$1993
to $\sim$1996. We will take an approximate epoch MJD 49718 ($\sim$1995.0)
in following discussion. There should be only a very small position offset
($<1''$) caused by pulsar proper motions, if any, over the NVSS observation
period,  much smaller than the position uncertainties of the NVSS sources
listed in Tab.~1. If the position of a pulsar was measured at an epoch
later than MJD 47000, we will not consider its proper motion during
the identification process for the same reason.

\input tab1.tex

The first step for identification is to check the position offset $\Delta$.
At this stage, we ignored the proper motion. If $\Delta$ is smaller than twice
of the total position uncertainty, ie. 
$\Delta \la 2 \sqrt{\sigma_{nvss}^2+\sigma_{psrcat}^2}$, 
then we attribute the NVSS source as being a positive identification of a
pulsar. This process yielded the first 90 positive detections. If any pulsar
position was obtained at an epoch several years ago, the pulsar must have
had only a very small proper motion so that the position offsets are not
significant.

Now we consider the remaining 16 sources more carefully, which are marked
with ``?'' in Column (14) of Tab.~1.

{\bf Nine Confusion cases:}
(a). PSR B0531+21 (Crab) and PSR B1951+32 are confused by their associated 
supernova remnants. We marked them in Notes, i.e. Column (14), of Tab.~1 with
``SNR''. 
(b). PSRs B1112+50, B1829$-$10 and B1831$-$00 are confused by their nearby
strong sources which have much larger flux density (more than 10 times) than
that from the pulsar catalog.
One NVSS source was detected $28.1''$ (formally $7.8\sigma$) away from
PSR B1920+21, too large to be proper motion for this distant pulsar 
(distance $\sim$12.5 kpc). We consider these detections unlikely and
mark with ``no'' in Notes to stand for ``no detection''.
(c). PSRs B1744-24A and J2129+1210A, (maybe also B1745-20 as indicated by Kaplan
et al. \cite{kapet98}), are confused by other continuum sources in the
host globular clusters Terzan-5 
and M15, (and NGC6440?), respectively. They are marked with ``glbc''.
(d). PSR B1718-35 is a marginal case, maybe confused by a source $19''.4$ away,
with 4.7$\sigma$ for position offset and 27.7 mJy in flux (pulsar: 10.0 mJy).\\

{\bf Seven Detection cases:}
(a). The position offset to PSR B1831$-$04 is only $4.75''$ (formally 
$2.3\sigma$, or $2.6\sigma$ rather than $14\sigma$ using the new position in
Kaplan et al. 1998), much smaller than the beam size of the NVSS. 
Although Kaplan et al. (1998) suggested otherwise, we believe the pulsar
is detected. The consistent flux densities of the pulsar and the NVSS 
source confirm the identification. We mark such a case as "yes" in the
Notes.
(b). PSRs B0823+26, B1133+16, B2016+28, (and B2154+40) have small
position offsets caused by proper motions
(see Sect.~3.3).
(c). PSR B1820-31 is detected with a position offset of
$12.5'' = 2.6\sigma$, as confirmed by consistent flux density, and more
importantly, by the highly linear polarization of the NVSS source.
(d). A marginal case is the strong pulsar PSR B2020+28. The NVSS survey detected
a very weak source $2.2\sigma$ away, too weak to believe the identification
(see more discussion below). However, highly linear polarization of the source
suggests that it is the pulsar. We mark in the Notes "yes?" for this case.

In all, the NVSS detected 97 pulsars, including the 73 which appeared in Kaplan
et al. (1998) and 24 new identifications.\footnote{
The PSR J1615-39 in Table 1 of Kaplan et al. (1998) is
missing from the pulsar catalog available to us.
}

\section{Discussion}
\begin{figure}
\psfig{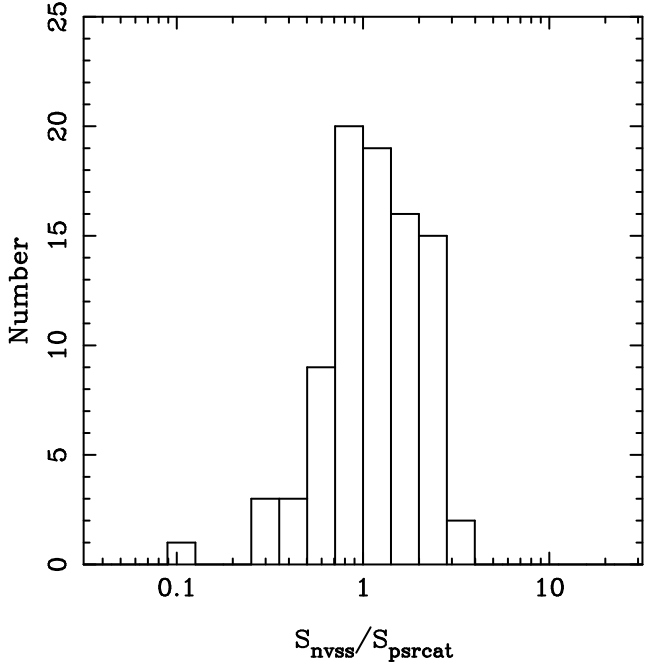}
\caption{The histograph for the flux comparison. Some pulsars (eg. these in
Table 1) have been missed in the left-half of the distribution (ie. when 
$S_{\rm nvss}/S_{\rm psrcat}<1$), because scintillation makes them
weaker than the NVSS sensitivity.}
\end{figure}
\subsection{Scintillation and undetected pulsars}

The VLA measurements of the flux densities 
$S_{1.4}$ of most identified pulsars, averaged
over about 84 MHz bandwidth and 3$\times$23 sec in time, are comparable
to the flux densities published in Lorimer et al. (\cite{loret95}) and Gould \& Lyne 
(\cite{gl98}). They are generally within a factor of 2 of the
published densities (see Fig.~1), but sometimes
up to a factor of 3 or more. Most of undetected  pulsars ($\sim$400)
have flux densities below 2 or 3 mJy. Interstellar scintillation (eg. Gupta
et al. \cite{gupet94}) both helps and hinders the detections (Cordes \& Lazio
\cite{cl91}). Some pulsars which have a flux density less than 2 mJy 
in the pulsar catalog have been detected
in the NVSS with a larger flux density. The scintillation effect is more
obvious for strong pulsars. For example, PSR B2020+28 should be as strong as
38.0 mJy, but in the NVSS it appears to be a highly polarized source 
of $3.6\pm0.5$
mJy. Among 61 pulsars with known flux densities
 larger than 5 mJy, about one fourth
were missed by the NVSS (as listed in Table 2), some due to scintillation,
some due to confusion (J. Condon, private communication).
\begin{table}  
\caption{Pulsars stronger than 5 mJy but not detected by the NVSS}
\begin{tabular}{cllll}
\hline 
  PSR B    & RA(2000)    & Dec(2000) & S$_{\rm 1.4}$ & Notes\\
           & !h~m~!s    & !~!$^o$~!$'$~!$''$ & mJy   &  \\ 

 1937$+$21 & 19~39~38.55 & $+$21~34~59.1 & 16.0 & conf.\\
 1800$-$21 & 18~03~51.35 & $-$21~37~07.2 & 14.6 & scin.\\
 2319$+$60 & 23~21~55.19 & $+$60~24~30.6 & 12   & scin.\\
 1845$-$01 & 18~48~24.00 & $-$01~23~58.2 & 10   & scin.\\
 2255$+$58 & 22~57~57.70 & $+$59~09~14.9 & !9.2 & scin.\\
 1839$-$04 & 18~42~26.49 & $-$03~59~59.2 & !8.5 & conf.\\
 1952$+$29 & 19~54~22.58 & $+$29~23~17.9 & !8   & scin.\\
 1815$-$14 & 18~18~23.79 & $-$14~22~35.9 & !7.4 & scin.\\
 1754$-$24 & 17~57~41.02 & $-$24~21~56.8 & !7.1 & conf.\\
 2011$+$38 & 20~13~10.49 & $+$38~45~44.8 & !6.4 & scin.\\
 1737$-$30 & 17~40~33.73 & $-$30~15~41.9 & !6   & scin.\\
 1919$+$21 & 19~21~44.80 & $+$21~53~01.8 & !6   & scin.\\
 1758$-$23 & 18~01~19.86 & $-$23~06~16.8 & !5.7 & scin.\\
 1849$+$00 & 18~52~28.00 & $+$00~31~55.9 & !5.2 & scin.\\
\hline
\end{tabular}
\end{table}

\subsection{Polarization}

When pulsars are observed as continuum radio sources, the polarized
intensity, $L$, and polarization position angle, $PA$, are calculated
from the integrated $Q$ and $U$ values of the final images, i.e., over
all the observation time and the bandwidth, so that 
 $$ L_{\rm nvss} = \sqrt{(\int_t Q)^2 + (\int_t U)^2}, \eqno(1)$$
and
 $$PA_{\rm nvss} = \frac{1}{2}\; \frac{180}{\pi}\;
    \arctan(\frac{\int_t U}{\int_t Q}). \eqno(2)$$
In pulsar observations, however, the total linearly polarized intensity
is
 $$ L_{\rm psr} \equiv \int_t \sqrt{Q^2+U^2}, \eqno(3)$$
and the polarization position angle PA is
 $$PA_{\rm psr} = \frac{1}{2}\; \frac{180}{\pi}\; \arctan(\frac{U}{Q})
 \eqno(4)$$
for each pulse longitude. The PA often swings more than $90\degr$ over
a pulse. Since a positive value of $Q$ or $U$ in one part of a pulse may
cancel a negative value in another part, it is believed that the pulsar
emission is depolarized in contiuum observations. Furthermore, the bandwidth
depolarization occurs for pulsars with high rotation measures.
Therefore the $L/S$ in Table 1 should be taken as the lower limit of pulsar
polarization.

\begin{figure}
\psfig{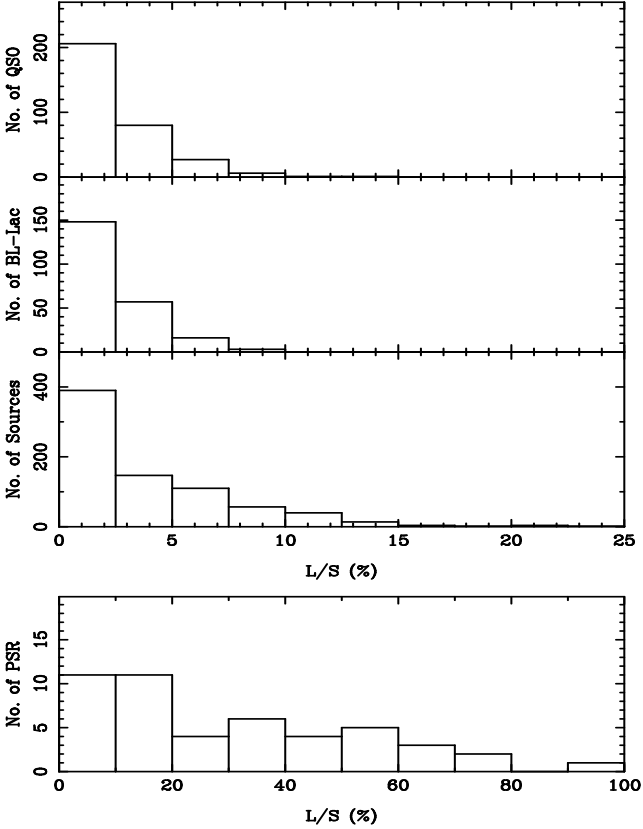}
\caption{Histograms of polarization percentage of a few kinds of objects:
quasars, BL-Lac objects, all radio sources in one sky area, and pulsars.
Note that the abscissa is up to 100\% for pulsars, but just 25\% for other
objects.}
\end{figure}
Even so, pulsars are still the sources with the highest polarization compared
to other kinds of objects (see Fig.~2). As seen from Table~1, some pulsars
have very high linear polarization, such as PSRs B1742-30 ($L/S \sim 90\%$)
and PSR B1929+10 ($L/S \sim 63\%$), even after the smearing and
depolarization.

Since the NVSS
has very accurate absolute position angle calibrations ($<0.2\degr$),
the well measured PA of a few pulsars (with error $\la 2\degr$) may
help to make an absolute PA calibration in pulsar observations.
One example is shown in Fig.~3.  First, using the VLA measurements of
PA at 1400 MHz and the RM values, we calculated the averaged PA over
the pulse at the observation frequency accordingly. Second, from the
pulsar observations, we got $PA$ for calibration pulsars using Eq.(2)
from the pulse profiles (including interpulse if applicable) of Stokes
parameters Q and U. Third we compared them to get an offset which
represents the instrument PA offset, and used it to calibrate all pulsar
observations.
\begin{figure}
\psfig{figure=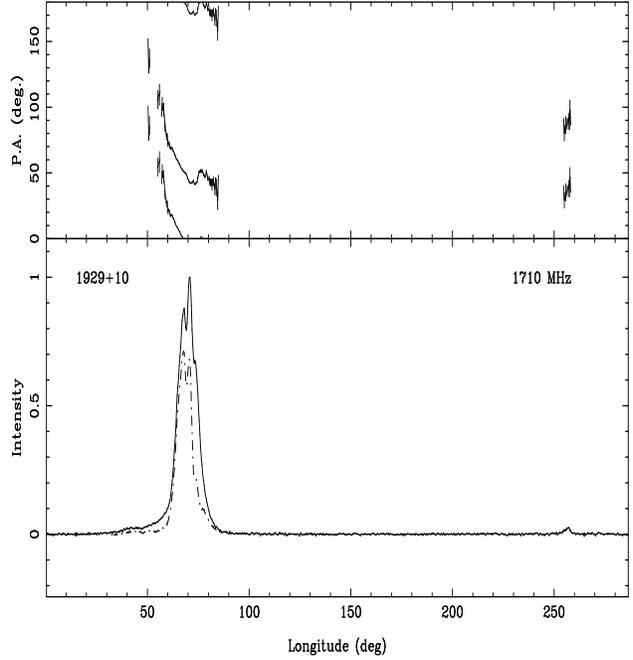,width=85mm,height=90mm,rotate=90}
\caption{Calibration for absolute polarization angle. Pulsar data were
observed by von Hoensbroech \& Xilouris (1997). In the lower panel,
the total intensity, $I$, and linearly polarized intensity, $L$, are plotted
with a thick continuum line and a dot-dash line, respectively. (The interpulse
is almost 100\% polarized.) In the top panel, the original PA data are
plotted with a thin line (and with an error bar on every second point),
and the calibrated PA data are plotted with a thick line.}
\end{figure}
\begin{table}  
\caption{Pulsar calibrators for absolute polarization angle}
\begin{tabular}{ccrr}
\hline 
 PSR J  &  PSR B  & \multicolumn{1}{c}{$PA_{\rm 1400MHz}^{\rm NVSS}$}&
 \multicolumn{1}{c}{$RM$} \\
        &         & \multicolumn{1}{c}{ ($\degr$) } & 
 \multicolumn{1}{c}{(rad~m$^{-2}$)} \\
\hline
1932$+$1059 & 1929$+$10 &$ 46\pm1 $&$ -6.1\pm1.0$ \\  
0742$-$2822 & 0740$-$28 &$-33\pm1 $&$150.4\pm0.1$ \\  
2022$+$5154 & 2021$+$51 &$ 11\pm2 $&$ -6.5\pm0.9$ \\  
0543$+$2329 & 0540$+$23 &$-25\pm2 $&$  8.7\pm0.7$ \\  
0630$-$2834 & 0628$-$28 &$-31\pm2 $&$ 46.2\pm0.1$ \\  
\hline
\end{tabular}
\end{table}

In Table 3, we listed 5 pulsars which can be used for calibration purposes.
All of them have strong linear polarized intensity that can be easily detected,
and their rotation measures $RM$ are either quite small  ($\la $10 rad~m$^{-2}$)
or accurately measured ($\sigma_{\rm RM} \la 1$ rad~m$^{-2}$). None of them
has any mode-changing (e.g. PSR B1237+25 and PSR B1822+09) or complicated
variations in PA across the profile (e.g. PSR B1933+16). All pulsars in
Table 3 satisfy $\sigma_{\rm PA} + \sigma_{\rm RM} \cdot \delta(\lambda^2)
< 3\degr$, where $\delta(\lambda^2)$ was the difference of the
wavelengths squared, and was taken as 1.0.

\subsection{Proper motions}
\begin{table}  
\begin{small}
\caption{Pulsar proper motions} 
\setlength{\tabcolsep}{1.1mm}
\begin{tabular}{crrcrr}
\hline 
  PSR B   & \multicolumn{1}{c}{$\Delta$ RA}  &
\multicolumn{1}{c}{$\Delta$ Dec }&Epoch& 
\multicolumn{1}{c}{$\mu_{\alpha}\cos\delta$}&
\multicolumn{1}{c}{$\mu_{\delta}$}\\ 
          &   \multicolumn{1}{c}{arcsec} & \multicolumn{1}{c}{arcsec} 
   & MJD & \multicolumn{1}{c}{mas yr$^{-1}$} 
& \multicolumn{1}{c}{mas yr$^{-1}$} \\ 
 \hline 
 0031$-$07& -2.53$\pm$1.84&  2.6$\pm$1.9&40690& -102$\pm$!74& -105$\pm$!78\\
 0329$+$54& -0.61$\pm$0.44&  0.9$\pm$0.6&40105&  -23$\pm$!16&   34$\pm$!23\\
 0450$-$18& -1.15$\pm$1.01& -2.3$\pm$1.0&46573& -133$\pm$117&  267$\pm$116\\
 0628$-$28&  1.07$\pm$1.21& -1.8$\pm$1.2&40585&   42$\pm$!48&   71$\pm$!48\\
 0740$-$28& -0.53$\pm$0.80& -1.3$\pm$0.8&46573&  -61$\pm$!92&  150$\pm$!93\\
 0823$+$26&  1.21$\pm$0.94& -3.7$\pm$1.0&40264&   46$\pm$!36& -142$\pm$!38\\
 0950$+$08& -0.15$\pm$0.45&  1.3$\pm$0.6&46058&  -14$\pm$!44&  129$\pm$!59\\
 1133$+$16& -0.29$\pm$0.72&  6.2$\pm$0.8&42364&  -14$\pm$!35&  307$\pm$!40\\
 1237$+$25& -0.54$\pm$0.82& -1.0$\pm$0.9&46460&  -61$\pm$!91& -112$\pm$100\\
 1541$+$09&  0.74$\pm$1.49& -2.0$\pm$1.5&42304&   36$\pm$!73&  -98$\pm$!74\\
 1749$-$28&  1.72$\pm$0.53& -0.8$\pm$0.9&40352&   67$\pm$!20&   31$\pm$!33\\
 1818$-$04& -6.74$\pm$2.69& -1.5$\pm$2.7&40614& -270$\pm$108&   60$\pm$108\\
 1857$-$26&  1.08$\pm$0.95&  1.7$\pm$1.1&46573&  125$\pm$110& -197$\pm$132\\
 1933$+$16& -0.14$\pm$0.58& -0.9$\pm$0.7&40213&   -5$\pm$!22&  -34$\pm$!26\\
 1946$+$35& -1.34$\pm$2.56& -3.1$\pm$2.9&42221&  -65$\pm$124& -151$\pm$141\\
 2016$+$28&  3.y42$\pm$0.79&  0.4$\pm$0.8&40105&  130$\pm$!30&   15$\pm$!30\\
 2021$+$51& -0.46$\pm$0.93& -0.6$\pm$1.0&40614&  -18$\pm$!37&  -24$\pm$!40\\
 2310$+$42& -2.75$\pm$1.98&  0.4$\pm$1.4&43891& -172$\pm$124&   25$\pm$!87\\
\hline
\end{tabular}
\end{small}
\end{table}

Pulsar proper motion is a very important quantity to be measured, so that
pulsar velociaties can be determined. Pulsar timing can be used to
determine the proper motions of millisecond pulsars because of their 
great timing stability (e.g. Nice \& Taylor \cite{nt95}). However, for
most pulsars, the proper motions can only be measured by determining the
pulsar position precisely at two or more well-seperated epochs using
interferometry (e.g. Fomalont et al. \cite{fomet97}). 

We compared the pulsar positions given in the pulsar catalog with those from
the NVSS whose epoch is simply taken as MJD=49718, and calculated
pulsar proper motions if possible. The results are listed in Table 4.
Pulsars with  uncertainties of  proper motion larger than 200
mas yr$^{-1}$ have been deleted. Because of the large uncertainty
of the NVSS positions, we obtained only a few significant measurements:
proper motion in declination direction of PSR B1133+16, and that in
right ascension of PSRs B0823+26 and B2016+28. While the former two
are consistent with the previous measurements made by Lyne, Anderson \&
Salter 
(\cite{las82}), the latter one is marginally not. Cross-checking with
Table 2 of Taylor, Manchester \& Lyne (\cite{tml93}), we found that
all other measurements in Table 4 are consistent with (though poorer
 than)
those given in the pulsar catalog, except for one new upper limit of
PSR B0031$-$07.
VLA A-array observations of these pulsars in Table 1 should provide
much more accurate positions, and hence could produce the first meaurement
of the proper motions of about 20 pulsars. 

PSR B0031$-$07 is a nearby pulsar with distance 0.68 kpc. Its proper
motion upper limit indicates that the pulsar has a velocity of 
$470\pm346$ km~s$^{-1}$, quite normal according to the pulsar velocity
distribution (Lyne \& Lorimar \cite{ll94}).

\section{Summary}
We identified about 97 strong pulsars from the NVSS catalog and 
presented the flux densities at 1.4 GHz. The parameters of linear
polarization are independent, but slightly different (see Eqs.(1),
and (2) above), measurements from those obtained from normal pulsar
observations.
Interstellar scintillation both helps and hinders the detection of pulsars.
Table 1 presents all known pulsars detected by the NVSS.
Well-calibrated VLA measurements of the average polarization angles
of 5 strong pulsars can be used for PA calibrations for pulsar
observations. By comparing the pulsar positions from the pulsar catalog
and those from the NVSS, we got a proper motion upper limit of PSR
B0031$-$07.

\begin{acknowledgements}
We thank the anonymous referee for his constructive suggestions which helped
to revise the paper significantly, and Drs. Dunc Lorimer, Elly Berkhuijsen,
R. Wielebinski and Paul Arendt for their helpful comments.
JLH is grateful for the hospitality of Prof. R. Wielebinski and Dr. R. Beck
during his stay at the MPIfR, Bonn as an exchange scholar between the
Chinese Academy of Sciences (CAS) and Max-Planck-Gesellschaft between
1997 May and 1998 August.
He also thanks the National Natural Science Foundation of China and the
Astronomical Committee of the CAS for continuous support.
\end{acknowledgements}

\end{document}

%% file: tab1.tex
\begin{table*}%
\vspace{240mm}
\begin{scriptsize}
  \begin{center}
    \leavevmode
\begin{rotate}{90}
\begin{tabular}{llcrcrccrrrrrl}
\multicolumn{14}{l}{\small {\bf Table 1:} NVSS sources around
pulsar positions}\\[1mm]
\hline
\hline
\multicolumn{1}{c}{(1)} & \multicolumn{1}{c}{(2)} & \multicolumn{1}{c}{(3)}&
\multicolumn{1}{c}{ (4)} & \multicolumn{1}{c}{(5)} &\multicolumn{1}{c}{(6)}& 
\multicolumn{1}{c}{(7)} & \multicolumn{1}{c}{(8)} & \multicolumn{1}{c}{(9)} & 
\multicolumn{1}{c}{(10)} & \multicolumn{1}{c}{(11)}&\multicolumn{1}{c}{(12)}&
\multicolumn{1}{c}{(13)} & \multicolumn{1}{c}{(14)} 
\\
     PSR J  &   PSR B  & RA(2000)  &  \multicolumn{1}{c}{ Dec(2000) }
& Epoch   & $S_{\rm 1.4}$ &  RA(2000)nvss      & Dec(2000)nvss   
& \multicolumn{1}{c}{$\Delta$} &$S_{1.4}\pm\sigma$&$L_{1.4}\pm\sigma$
&  \multicolumn{1}{c}{$L/S$} & \multicolumn{1}{c}{PA$\pm\sigma$} & Notes
\\
            &          & \multicolumn{1}{l}{~h ~m !!s !!$\sigma$} & $^o$! $'$! $''$!! $\sigma$ &
MJD  & mJy   &\multicolumn{1}{l}{~h ~m !s !!!$\sigma$} & !$^o$ !$'$ !$''$!
!$\sigma$  & ($''$) & \multicolumn{1}{c}{(mJy)} & \multicolumn{1}{c}{(mJy)}
& \multicolumn{1}{c}{(\%)}     & \multicolumn{1}{c}{(deg)}        &       
\\
\hline
0014$+$4746 &0011$+$47 &00 14 17.74 .06&   47 46 33.1 0.6& 48416 & 3.0& 00 14 18.18 0.44&$+$47 46 39.7 4.5& 7.97&  4.9$\pm$0.5&  1.95$\pm$0.60& 40$\pm$13&   24$\pm$ 6&     \\  
0034$-$0721 &0031$-$07 &00 34 08.88 .03&$-$07 21 53.4 0.7& 40690 &11.0& 00 34 08.71 0.12&$-$07 21 56.0 1.8& 3.66& 14.6$\pm$1.1&  0.89$\pm$0.61&  6$\pm$ 4&$-$11$\pm$14&     \\
0055$+$5117 &0052$+$51 &00 55 45.39 .03&   51 17 24.8 0.3& 46116 & 1.5& 00 55 45.12 1.19&$+$51 17 22.9 5.1& 3.19&  4.1$\pm$0.4&  1.34$\pm$0.83& 33$\pm$20&    0$\pm$10&     \\
0139$+$5814 &0136$+$57 &01 39 19.77 .00&   58 14 31.8 0.0& 48382 & 4.6& 01 39 19.99 0.46&$+$58 14 37.4 5.0& 5.90&  4.0$\pm$0.4&  2.09$\pm$0.56& 52$\pm$15&$-$22$\pm$ 5&     \\ 
0141$+$6009 &0138$+$59 &01 41 39.95 .00&   60 09 32.3 0.0& 46573 & 4.5& 01 41 39.73 0.41&$+$60 09 26.5 3.2& 6.06&  5.4$\pm$0.4&  1.11$\pm$0.57& 21$\pm$11& $-$3$\pm$10& *        
\\[1mm]
0304$+$1932 &0301$+$19 &03 04 33.11 .01&   19 32 50.7 0.1& 46058 & 3.0& 03 04 33.04 0.15&$+$19 32 47.6 2.2& 3.20&  6.4$\pm$0.5&  0.42$\pm$0.38&  7$\pm$ 6&   ......   &     \\
0332$+$5434 &0329$+$54 &03 32 59.35 .01&   54 34 43.2 0.1& 40105 & 203& 03 32 59.28 0.05&$+$54 34 44.1 0.6& 1.12&150.7$\pm$5.4&  5.52$\pm$0.53&  4$\pm$ 0&   44$\pm$ 2&     \\ 
0357$+$5236 &0353$+$52 &03 57 44.82 .00&   52 36 57.7 0.1& 48416 & 1.9& 03 57 44.53 0.68&$+$52 36 51.5 7.9& 6.72&  3.2$\pm$0.5&  1.08$\pm$0.69& 34$\pm$22&   ......   &     \\
0358$+$5413 &0355$+$54 &03 58 53.70 .00&   54 13 13.6 0.0& 46573 &23.0& 03 58 53.89 0.16&$+$54 13 15.9 1.6& 2.85& 10.3$\pm$0.5&  3.57$\pm$0.49& 35$\pm$ 5&   19$\pm$ 3&     \\
0406$+$6138 &0402$+$61 &04 06 30.05 .01&   61 38 40.7 0.2& 48416 & 2.8& 04 06 30.27 1.27&$+$61 38 30.7 10.& 10.1&  3.7$\pm$0.4&  0.11$\pm$0.91&  3$\pm$25&   ......   &       
\\[1mm]
0452$-$1759 &0450$-$18 &04 52 34.10 .00&$-$17 59 23.5 0.1& 46573 & 5.3& 04 52 34.02 0.07&$-$17 59 21.2 1.0& 2.56& 14.5$\pm$0.6&  1.76$\pm$0.40& 12$\pm$ 3&$-$75$\pm$ 5& *   \\
0454$+$5543 &0450$+$55 &04 54 07.62 .00&   55 43 41.2 0.1& 46460 &13.0& 04 54 07.47 0.22&$+$55 43 41.4 2.3& 1.25&  7.8$\pm$0.5&  1.75$\pm$0.47& 22$\pm$ 6&$-$54$\pm$ 7&     \\
0528$+$2200 &0525$+$21 &05 28 52.34 .02&   22 00 00.2 5.0& 41994 & 9.0& 05 28 52.26 0.65&$+$22 00 06.8 13.& 6.72&  2.8$\pm$0.5&  1.53$\pm$0.96&   ...... &   19$\pm$12&     \\
0534$+$2200 &0531$+$21 &05 34 31.97 .01&   22 00 52.1 0.1& 40675 &14.0& 05 34 31.56 0.03&$+$22 00 43.2 0.6& 21.5& 4106$\pm$123&  99.9$\pm$0.54&  ......  &   90$\pm$ 0&?:SNR\\
0538$+$2817 & ......   &05 38 25.06 .04&   28 17 11.0 5.0& 49444 &    & 05 38 25.11 0.29&$+$28 17 17.7 4.4& 6.73&  4.2$\pm$0.5&  2.95$\pm$0.62& 70$\pm$17&$-$58$\pm$ 5&    
\\[1mm]
0543$+$2329 &0540$+$23 &05 43 09.65 .00&   23 29 06.1 0.0& 47555 & 9.0& 05 43 09.66 0.11&$+$23 29 04.1 1.7& 1.93& 15.8$\pm$1.4&  7.38$\pm$0.72& 47$\pm$ 6&$-$25$\pm$ 2&     \\
0629$+$2415 &0626$+$24 &06 29 05.72 .00&   24 15 41.6 0.1& 47555 & 3.2& 06 29 05.74 0.23&$+$24 15 38.3 3.2& 3.34&  4.7$\pm$0.5&               &          &            &     \\
0630$-$2834 &0628$-$28 &06 30 49.53 .01&$-$28 34 43.6 0.1& 40585 &23.0& 06 30 49.61 0.09&$-$28 34 41.8 1.2& 2.04& 13.8$\pm$0.6&  4.01$\pm$0.44& 29$\pm$ 3&$-$31$\pm$ 2&     \\
0742$-$2822 &0740$-$28 &07 42 49.07 .00&$-$28 22 44.0 0.1& 46573 &23.0& 07 42 49.03 0.06&$-$28 22 42.7 0.8& 1.46& 22.3$\pm$0.8& 13.44$\pm$0.45& 60$\pm$ 3&$-$33$\pm$ 1&     \\
0814$+$7429 &0809$+$74 &08 14 59.44 .04&   74 29 05.8 0.1& 48382 &10.0& 08 14 58.88 0.62&$+$74 29 05.7 2.2& 2.23&  7.7$\pm$0.5&  2.65$\pm$0.50& 34$\pm$ 7&$-$45$\pm$ 4&      
\\[1mm]
0820$-$1350 &0818$-$13 &08 20 26.36 .01&$-$13 50 55.2 0.1& 46573 & 7.0& 08 20 26.38 0.21&$-$13 50 53.4 3.1& 1.80&  5.5$\pm$0.5&  0.29$\pm$0.45&  5$\pm$ 8&   ......   &     \\
0826$+$2637 &0823$+$26 &08 26 51.31 .00&   26 37 25.6 0.1& 40264 &10.0& 08 26 51.40 0.07&$+$26 37 21.9 1.0& 3.88& 17.1$\pm$0.7&  2.24$\pm$0.44& 13$\pm$ 3&   75$\pm$ 4&?:yes\\
0837$+$0610 &0834$+$06 &08 37 05.65 .00&   06 10 14.1 0.1& 46058 & 4.0& 08 37 04.24 0.39&$+$06 10 13.4 8.2& 21.0&  3.7$\pm$0.5&  0.52$\pm$0.64& 14$\pm$17&   ......   &     \\
0846$-$3533 &0844$-$35 &08 46 05.86 .14&$-$35 33 39.9 1.5& 43557 & 4.0& 08 46 05.64 0.37&$-$35 33 31.2 5.8& 9.12&  3.2$\pm$0.4&  0.39$\pm$0.50& 12$\pm$16&   ......   &     \\
0922$+$0638 &0919$+$06 &09 22 13.98 .00&   06 38 21.6 0.0& 46573 & 4.2& 09 22 14.39 0.17&$+$06 38 24.3 2.5& 6.69&  6.9$\pm$0.5&  3.81$\pm$0.47& 55$\pm$ 8&   41$\pm$ 2&      
\\[1mm]
0953$+$0755 &0950$+$08 &09 53 09.32 .00&   07 55 35.6 0.0& 46058 &84.0& 09 53 09.31 0.03&$+$07 55 36.9 0.6& 1.30& 92.0$\pm$2.8&  2.78$\pm$0.42&  3$\pm$ 0&$-$14$\pm$ 4&     \\
1012$+$5307 & ......   &10 12 33.43 .00&   53 07 02.6 0.0& 49220 & 2.8& 10 12 33.31 0.36&$+$53 07 04.9 4.8& 2.58&  4.5$\pm$0.4&  1.95$\pm$0.48& 43$\pm$11&$-$17$\pm$ 5&     \\
1022$+$1001 & ......   &10 22 58.05 .06&   10 01 54.0 3.0& 49780 & 2.3& 10 22 57.92 0.34&$+$10 01 53.5 9.2& 2.00&  3.5$\pm$0.4&  1.46$\pm$0.61& 42$\pm$18&    9$\pm$ 8&     \\
1034$-$3224 & ......   &10 34 19.48 .01&$-$32 24 26.3 0.1& 49020 & 4.7& 10 34 19.18 0.26&$-$32 24 23.3 4.2& 4.83&  4.4$\pm$0.5&      ......   &  ......  &  ......    & *   \\
1115$+$5030 &1112$+$50 &11 15 38.35 .02&   50 30 13.5 0.3& 44240 & 3.0& 11 15 38.56 0.06&$+$50 30 25.6 0.7& 12.2& 35.2$\pm$1.1&  0.19$\pm$0.38&  1$\pm$ 1&  ......    &?:no    
\\[1mm]
1136$+$1551 &1133$+$16 &11 36 03.30 .00&   15 51 00.7 0.1& 42364 &32.0& 11 36 03.28 0.05&$+$15 51 06.9 0.8& 6.16& 21.2$\pm$0.8&  1.54$\pm$0.39&  7$\pm$ 2&$-$57$\pm$ 6&?:yes\\
1239$+$2453 &1237$+$25 &12 39 40.47 .00&   24 53 49.3 0.0& 46460 &10.0& 12 39 40.43 0.06&$+$24 53 48.3 0.9& 1.16& 20.5$\pm$0.7&  7.49$\pm$0.38& 37$\pm$ 2&   84$\pm$ 1& mode\\
1320$-$3512 & ......   &13 20 12.70 .10&$-$35 12 24.0 2.0& 48734 &    & 13 20 12.58 0.15&$-$35 12 24.1 2.0& 1.51&  7.9$\pm$0.5&  1.28$\pm$0.47& 16$\pm$ 6&   60$\pm$ 7& *   \\
1509$+$5531 &1508$+$55 &15 09 25.72 .01&   55 31 33.0 0.1& 48383 & 8.0& 15 09 25.78 0.23&$+$55 31 34.6 2.0& 1.74&  7.7$\pm$0.5&  1.03$\pm$0.40& 13$\pm$ 5&    8$\pm$ 9&     \\
1518$+$4904 & ......   &15 18 16.60 .10&   49 04 35.0 1.0&       &    & 15 18 17.08 0.42&$+$49 04 30.9 4.6& 6.26&  4.4$\pm$0.4&  0.12$\pm$0.48&  3$\pm$11&   ......   &       
\\[1mm] 
1543$+$0929 &1541$+$09 &15 43 38.83 .01&   09 29 16.8 0.2& 42304 & 5.9& 15 43 38.88 0.10&$+$09 29 14.8 1.5& 2.10& 10.0$\pm$0.5&  1.72$\pm$0.47& 17$\pm$ 5&   64$\pm$ 6& *   \\
1543$-$0620 &1540$-$06 &15 43 30.17 .01&$-$06 20 45.3 0.1& 46573 & 2.0& 15 43 30.33 0.43&$-$06 20 46.8 6.8& 2.84&  2.8$\pm$0.5&    ......     &   ...... &  ......    & *   \\ 
1603$-$2531 &  ......  &16 03 04.88 .00&$-$25 31 48.3 0.1& 49408 &    & 16 03 04.97 0.61&$-$25 31 47.9 6.8& 1.21&  3.1$\pm$0.5&  0.80$\pm$0.81& 26$\pm$26&  ......    & *   \\
1607$-$0032 &1604$-$00 &16 07 12.12 .00&$-$00 32 40.2 0.1& 42307 & 5.0& 16 07 12.00 0.27&$-$00 32 48.8 5.7& 8.73&  4.1$\pm$0.5&    ......     &  ......  &  ......    &     \\
1643$-$1224 &  ......  &16 43 38.15 .00&$-$12 24 58.7 0.0& 49524 & 3.1& 16 43 38.60 0.25&$-$12 24 51.8 3.7& 9.50&  3.9$\pm$0.4&  0.10$\pm$0.47&  3$\pm$12&  ......    &     
\\[1mm]
1645$-$0317 &1642$-$03 &16 45 02.03 .00&$-$03 17 58.3 0.1& 40414 &21.0& 16 45 02.34 0.40&$-$03 18 10.5 7.1& 13.1&  8.3$\pm$1.5&    .......     &   ...... & ......    &  *  \\
1700$-$3312 &  ......  &17 00 53.02 .01&$-$33 12 45.1 0.6& 49424 &    & 17 00 52.52 0.47&$-$33 12 59.8 6.7& 16.0&  2.8$\pm$0.5&  0.72$\pm$0.77& 26$\pm$28&  ......    &  *  \\
1703$-$3241 &1700$-$32 &17 03 22.37 .12&$-$32 41 45.2 3.5& 48000 & 6.0& 17 03 22.34 0.25&$-$32 41 51.6 2.7& 6.45&  6.7$\pm$0.5&  1.23$\pm$0.85& 18$\pm$13&  ......    &     \\
1705$-$1906 &1702$-$19 &17 05 36.11 .01&$-$19 06 38.5 0.7& 48331 & 8.0& 17 05 35.83 0.19&$-$19 06 38.9 3.9& 3.98&  5.6$\pm$0.5&  2.82$\pm$0.68& 50$\pm$13&$-$39$\pm$ 7&     \\
1705$-$3423 &  ......  &17 05 42.37 .00&$-$34 23 44.5 0.2& 49500 &    & 17 05 42.30 0.38&$-$34 23 41.2 5.6& 3.40&  4.2$\pm$0.5&  2.14$\pm$0.67& 51$\pm$17&$-$85$\pm$15& *  
\\[1mm]
1708$-$3426 &  ......  &17 08 57.75 .01&$-$34 26 47.5 0.8& 49339 & 2.4& 17 08 58.33 0.47&$-$34 26 33.7 11.& 15.5&  2.9$\pm$0.5&  1.42$\pm$0.74& 49$\pm$27&$-$23$\pm$14& *   \\
1713$+$0747 &  ......  &17 13 49.52 .00&   07 47 37.5 0.0& 49056 & 3.0& 17 13 50.11 0.49&$+$07 47 38.4 3.9& 8.81&  8.0$\pm$1.4&     ......    &   ...... &  ......    & *   \\
1721$-$3532 &1718$-$35 &17 21 32.80 .02&$-$35 32 46.6 0.9& 48379 &10.0& 17 21 34.04 0.10&$-$35 32 58.5 3.7& 19.2& 27.7$\pm$2.1&  2.42$\pm$0.99&  9$\pm$ 4&   13$\pm$20&?:no?\\
1733$-$2228 &1730$-$22 &17 33 26.41 .06&$-$22 28 36.4 16.& 48417 & 2.2& 17 33 26.60 0.30&$-$22 28 39.9 4.1& 4.37&  3.2$\pm$0.4&  0.27$\pm$0.59& 8$\pm$18&  ......     &     \\
1740$+$1311 &1737$+$13 &17 40 07.37 .02&   13 11 57.5 0.3& 43893 & 3.9& 17 40 07.29 0.46&$+$13 11 57.9 7.2& 1.29&  3.8$\pm$0.5&  0.47$\pm$0.86& 12$\pm$23&  ......    &   
\\[1mm]
1741$-$0840 &1738$-$08 &17 41 22.54 .03&$-$08 40 32.7 1.6& 48417 & 1.4& 17 41 22.48 0.32&$-$08 40 29.7 6.3& 3.11&  3.3$\pm$0.5&  0.70$\pm$0.61& 21$\pm$19&   ......   & *   \\
1741$-$3927 &1737$-$39 &17 41 18.04 .03&$-$39 27 37.8 1.4& 43558 & 3.5& 17 41 17.42 0.31&$-$39 27 45.0 4.7& 10.2&  4.5$\pm$0.5&      ......   &  ......  &    ......  &     \\
1744$-$1134 &  ......  &17 44 29.39 .00&$-$11 34 54.6 0.0& 49882 & 1.0& 17 44 29.31 0.30&$-$11 34 58.6 5.5& 4.18&  3.8$\pm$0.5&  2.99$\pm$0.75& 79$\pm$22&   48$\pm$ 5& *   \\ 
1745$-$3040 &1742$-$30 &17 45 56.30 .00&$-$30 40 23.6 0.3& 46956 &14.0& 17 45 56.28 0.17&$-$30 40 28.6 2.5& 5.02&  6.6$\pm$0.5&  5.96$\pm$0.44& 90$\pm$10&$-$38$\pm$ 2&     \\
1748$-$1300 &1745$-$12 &17 48 17.37 .04&$-$13 00 53.3 2.0& 43891 & 2.0& 17 48 17.54 0.30&$-$13 00 39.6 8.6& 14.0&  4.4$\pm$0.5&   ......      &  ......  &   ......   &     
\\[1mm]
\hline 
\end{tabular}
\end{rotate} 
\end{center}
\end{scriptsize}
\end{table*} 

\begin{table*}
\vspace{240mm}
\begin{scriptsize}
  \begin{center}
    \leavevmode
\begin{rotate}{90}
\begin{tabular}{llcrcrccrrrrrl}
\multicolumn{14}{l}{\small {\bf Table 1:} --- Continued
}\\[1mm]
\hline
\hline
\multicolumn{1}{c}{(1)} & \multicolumn{1}{c}{(2)} & \multicolumn{1}{c}{(3)}&
\multicolumn{1}{c}{ (4)} & \multicolumn{1}{c}{(5)} &\multicolumn{1}{c}{(6)}& 
\multicolumn{1}{c}{(7)} & \multicolumn{1}{c}{(8)} & \multicolumn{1}{c}{(9)} & 
\multicolumn{1}{c}{(10)} & \multicolumn{1}{c}{(11)}&\multicolumn{1}{c}{(12)}&
\multicolumn{1}{c}{(13)} & \multicolumn{1}{c}{(14)} 
\\
     PSR J  &   PSR B  & RA(2000)  &  \multicolumn{1}{c}{ Dec(2000) }
& Epoch   & $S_{\rm 1.4}$ &  RA(2000)nvss      & Dec(2000)nvss   
& \multicolumn{1}{c}{$\Delta$} &$S_{1.4}\pm\sigma$&$L_{1.4}\pm\sigma$
&  \multicolumn{1}{c}{$L/S$} & \multicolumn{1}{c}{PA$\pm\sigma$} & Notes$^*$
\\
            &          & \multicolumn{1}{l}{~h ~m !!s !!$\sigma$} & $^o$! $'$! $''$!! $\sigma$ &
MJD  & mJy   &\multicolumn{1}{l}{~h ~m !s !!!$\sigma$} & !$^o$ !$'$ !$''$!
!$\sigma$  & ($''$) & \multicolumn{1}{c}{(mJy)} & \multicolumn{1}{c}{(mJy)}
& \multicolumn{1}{c}{(\%)}     & \multicolumn{1}{c}{(deg)}        &       
\\
\hline
1748$-$2021 &1745$-$20 &17 48 52.61 .10&$-$20 21 40.1 3.0& 49215 & 1.5& 17 48 53.14 0.43&$-$20 21 34.6 9.3& 9.24&  4.5$\pm$0.5&  0.74$\pm$1.14& 16$\pm$25&  ......    &(glbc) *\\
1748$-$2446A&1744$-$24A&17 48 02.25 .00&$-$24 46 37.7 0.5& 48270 & 2.5& 17 48 04.37 0.39&$-$24 46 41.5 4.6& 29.1&  3.7$\pm$0.5&  2.24$\pm$0.65& 61$\pm$19&$-$84$\pm$ 7&?:glbc\\
1750$-$3157 &1747$-$31 &17 50 47.29 .03&$-$31 57 41.3 3.0& 48379 & 1.4& 17 50 47.96 0.59&$-$31 58 01.3 10.& 21.7&  3.5$\pm$0.4&  1.78$\pm$1.06& 51$\pm$31&$-$78$\pm$24&  *  \\
1752$-$2806 &1749$-$28 &17 52 58.69 .00&$-$28 06 38.3 0.5& 40352 &36.0& 17 52 58.82 0.04&$-$28 06 37.5 0.7& 1.88& 41.6$\pm$1.3&  2.36$\pm$0.51&  6$\pm$ 1&$-$16$\pm$ 5&     \\
1801$-$2920 &1758$-$29 &18 01 46.83 .03&$-$29 20 37.1 3.0& 48377 & 2.8& 18 01 46.89 0.55&$-$29 20 33.0 7.2& 4.16&  2.2$\pm$0.5&  1.84$\pm$0.62& 84$\pm$34&$-$66$\pm$14&      
\\[1mm]
1807$-$0847 &1804$-$08 &18 07 38.02 .01&$-$08 47 43.1 0.2& 46573 &16.0& 18 07 38.08 0.12&$-$08 47 41.5 2.0& 1.83&  9.6$\pm$0.5&     ......    &  ......  &  ......    &     \\ 
1817$-$3618 &1813$-$36 &18 17 05.76 .02&$-$36 18 05.5 0.9& 48426 & 2.0& 18 17 05.44 0.44&$-$36 18 00.7 9.5& 6.13&  4.8$\pm$0.5&  0.87$\pm$1.30& 18$\pm$27&  ......    &     \\
1820$-$0427 &1818$-$04 &18 20 52.62 .00&$-$04 27 38.5 0.1& 40614 & 8.0& 18 20 52.17 0.18&$-$04 27 37.0 2.7& 6.87&  5.9$\pm$0.5&     ......    &  ......  &  ......    & *   \\
1822$-$2256 &1819$-$22 &18 22 58.96 .15&$-$22 56 49.3 54.& 48382 & 3.2& 18 22 58.68 0.51&$-$22 56 31.0 8.5& 18.7&  4.0$\pm$0.5&     ......    &  ......  &  ......    &     \\
1823$+$0550 &1821$+$05 &18 23 30.98 .03&   05 50 24.7 0.5& 44240 & 1.7& 18 23 31.17 0.33&$+$05 50 16.2 8.4& 8.97&  3.2$\pm$0.5&  0.45$\pm$0.76& 14$\pm$24&  ......    & *       
\\[1mm]
1823$-$3106 &1820$-$31 &18 23 46.78 .00&$-$31 06 49.7 0.2& 48382 & 2.5& 18 23 46.11 0.24&$-$31 06 40.7 3.7& 12.5&  4.9$\pm$0.5&  2.60$\pm$0.65& 53$\pm$14&  89$\pm$ 7 &?:yes\\
1824$-$1945 &1821$-$19 &18 24 00.45 .04&$-$19 45 50.7 8.0& 46800 & 6.1& 18 23 59.94 0.33&$-$19 45 49.1 3.6& 7.30&  5.1$\pm$0.5&     ......    &   ...... &    ......  &     \\
1825$-$0935 &1822$-$09 &18 25 30.60 .01&$-$09 35 22.8 0.4& 48381 &11.0& 18 25 30.60 0.05&$-$09 35 21.7 0.9& 1.10& 29.2$\pm$1.4&  9.43$\pm$0.66& 32$\pm$ 3&$-$39$\pm$ 1&     \\
1829$-$1751 &1826$-$17 &18 29 43.12 .01&$-$17 51 02.9 1.8& 46944 &10.3& 18 29 43.12 0.13&$-$17 51 03.8 2.6& 0.90& 12.2$\pm$1.2&    ......     &   ...... &   ......   &     \\
1832$-$1021 &1829$-$10 &18 32 40.90 .01&$-$10 21 34.2 0.5& 48000 & 0.6& 18 32 40.58 0.16&$-$10 21 19.1 4.7& 15.8& 11.7$\pm$1.4&     ......    &   ...... &   ......   &?:no 
\\[1mm]
1834$-$0010 &1831$-$00 &18 34 17.24 .04&$-$00 10 49.5 0.9& 46071 &    & 18 34 18.77 0.03&$-$00 10 55.2 0.6& 23.7& 70.6$\pm$2.2&  0.31$\pm$0.64&  ......  &     ...... &?:no \\ 
1834$-$0426 &1831$-$04 &18 34 25.62 .02&$-$04 26 15.4 0.6& 48000 &15.0& 18 34 25.48 0.08&$-$04 26 11.2 1.5& 4.75& 16.7$\pm$0.7&  0.23$\pm$0.59&  1$\pm$ 4&     ...... &?:yes\\ 
1836$-$1008 &1834$-$10 &18 36 53.89 .09&$-$10 08 09.3 5.0& 43893 & 5.0& 18 36 52.99 0.61&$-$10 08 13.0 18.& 13.8&  4.3$\pm$0.5&    ......     &  ......  &     ...... & *   \\
1840$+$5640 &1839$+$56 &18 40 44.59 .05&   56 40 55.5 0.4& 48381 & 4.0& 18 40 44.20 0.25&$+$56 40 55.8 2.3& 3.21& 10.0$\pm$1.0&  2.69$\pm$0.56& 27$\pm$ 6&  31$\pm$ 4 &     \\
1841$+$0912 &1839$+$09 &18 41 55.97 .01&   09 12 07.7 0.2& 43890 & 1.7& 18 41 56.35 0.46&$+$09 12 17.2 5.7& 11.1&  3.7$\pm$0.5&  1.85$\pm$0.74& 50$\pm$21&$-$21$\pm$13&   
\\[1mm]
1857$+$0943 &1855$+$09 &18 57 36.39 .00&   09 43 17.3 0.0& 47526 & 4.0& 18 57 36.27 0.35&$+$09 43 41.8 18.& 24.6&  3.8$\pm$0.5&     ......    &  ......  &    ......  &     \\
1900$-$2600 &1857$-$26 &19 00 47.60 .01&$-$26 00 43.1 0.3& 46573 &13.0& 19 00 47.68 0.07&$-$26 00 44.8 1.1& 2.02& 17.8$\pm$0.7&  1.99$\pm$0.60& 11$\pm$ 3& $-$6$\pm$ 5&     \\
1903$+$0135 &1900$+$01 &19 03 29.98 .00&   01 35 38.2 0.1& 42345 & 2.0& 19 03 30.43 0.74&$+$01 35 33.3 8.5& 8.28&  4.3$\pm$0.6&    ......     &   ...... &    ......  &     \\
1910$+$0358 &1907$+$03 &19 10 09.08 .20&   03 58 26.8 5.0& 43984 &    & 19 10 09.73 0.55&$+$03 58 33.7 7.8& 11.8&  4.7$\pm$0.5&   ......      &   ...... &    ......  &     \\
1913$-$0440 &1911$-$04 &19 13 54.18 .01&$-$04 40 47.6 0.4& 41902 & 4.4& 19 13 54.37 0.21&$-$04 40 46.9 4.4& 2.92&  5.1$\pm$0.5&  0.15$\pm$0.80&  3$\pm$16&    ......  &         
\\[1mm]
1916$+$1312 &1914$+$13 &19 16 58.69 .01&   13 12 49.7 0.3& 48382 & 1.9& 19 16 58.96 0.78&$+$13 12 49.4 16.& 4.02&  3.9$\pm$0.5&  2.25$\pm$1.40& 58$\pm$37&$-$35$\pm$15&     \\
1922$+$2110 &1920$+$21 &19 22 53.55 .00&   21 10 42.3 0.2& 42546 & 1.4& 19 22 52.23 0.18&$+$21 11 03.4 2.6& 28.1&  6.0$\pm$0.5&    ......     &   ...... &   ......   &?:no \\
1932$+$1059 &1929$+$10 &19 32 13.90 .00&   10 59 32.0 0.1& 48381 &41.0& 19 32 14.03 0.05&$+$10 59 31.9 0.8& 1.90& 27.1$\pm$0.9& 18.68$\pm$0.57& 69$\pm$ 3&  46$\pm$ 1 &     \\
1935$+$1616 &1933$+$16 &19 35 47.83 .00&   16 16 40.6 0.0& 40213 &42.0& 19 35 47.82 0.04&$+$16 16 39.7 0.7& 0.91& 43.3$\pm$1.4&  7.76$\pm$0.53& 18$\pm$ 1&$-$30$\pm$1 &     \\
1937$+$2544 &1935$+$25 &19 37 01.26 .01&   25 44 13.7 0.1& 48415 & 2.3& 19 37 01.39 0.42&$+$25 44 15.9 5.7& 2.75&  2.8$\pm$0.5&  0.63$\pm$0.51& 23$\pm$19&   ......   &       
\\[1mm]
1946$+$1805 &1944$+$17 &19 46 53.04 .00&   18 05 41.6 0.1& 42320 &10.0& 19 46 53.00 0.37&$+$18 05 43.2 4.1& 1.64&  3.5$\pm$0.4&  0.81$\pm$0.58& 23$\pm$17&  29$\pm$12 &     \\
1948$+$3540 &1946$+$35 &19 48 25.04 .00&   35 40 11.3 0.0& 42221 & 8.3& 19 48 24.93 0.21&$+$35 40 08.2 2.9& 3.44&  5.9$\pm$0.5&  1.40$\pm$0.57& 24$\pm$10&  81$\pm$11 &     \\
1952$+$3252 &1951$+$32 &19 52 58.30 .01&   32 52 40.4 0.2& 47005 & 1.0& 19 52 58.59 0.04&$+$32 52 43.3 0.6& 4.71& 979.$\pm$36.&  5.55$\pm$0.72&  1$\pm$ 0&  55$\pm$ 3 &?:SNR\\
2018$+$2839 &2016$+$28 &20 18 03.85 .00&   28 39 54.3 0.0& 40105 &30.0& 20 18 04.11 0.06&$+$28 39 54.7 0.8& 3.62& 24.7$\pm$0.8&  1.40$\pm$0.53&  6$\pm$ 2&$-$58$\pm$10&?:yes *\\
2022$+$2854 &2020$+$28 &20 22 37.08 .00&   28 54 23.5 0.0& 42370 &38.0& 20 22 38.95 0.72&$+$28 54 19.1 8.1& 24.9&  3.6$\pm$0.5&  2.21$\pm$1.01& 61$\pm$29&  57$\pm$11 &?:yes? *   
\\[1mm]
2022$+$5154 &2021$+$51 &20 22 49.90 .00&   51 54 50.1 0.0& 40614 &27.0& 20 22 49.85 0.10&$+$51 54 49.5 1.0& 0.74& 17.2$\pm$0.7&  6.10$\pm$0.46& 35$\pm$ 3&  11$\pm$ 2 &     \\
2023$+$5037 &2022$+$50 &20 23 41.94 .01&   50 37 34.8 0.0& 48591 & 2.2& 20 23 41.95 0.64&$+$50 37 39.9 5.7& 5.10&  7.4$\pm$1.4&  0.99$\pm$0.98& 13$\pm$13&   ......   &   * \\
2048$-$1616 &2045$-$16 &20 48 35.47 .00&$-$16 16 44.4 0.1& 46573 &13.0& 20 48 35.69 0.14&$-$16 16 45.4 2.5& 3.27&  7.0$\pm$0.5&  1.16$\pm$0.45& 17$\pm$ 7&  53$\pm$ 9 &     \\
2108$+$4441 &2106$+$44 &21 08 20.48 .01&   44 41 48.8 0.1& 48383 & 5.4& 21 08 20.60 0.33&$+$44 41 48.8 3.6& 1.30&  4.7$\pm$0.5&  0.27$\pm$0.44&  6$\pm$ 9&   ......   &     \\
2113$+$4644 &2111$+$46 &21 13 24.29 .01&   46 44 08.7 0.1& 48382 &19.0& 21 13 24.25 0.06&$+$46 44 08.6 0.7& 0.38& 37.6$\pm$1.5&  4.07$\pm$0.50& 11$\pm$ 1&$-$40$\pm$ 3&       
\\[1mm]
2124$-$3358 & ......   &21 24 43.86 .00&$-$33 58 44.2 0.0& 49674 & 1.6& 21 24 44.11 0.43&$-$33 58 35.2 11.& 9.56&  3.0$\pm$0.5&  0.14$\pm$0.68&  4$\pm$23&   ......   &     \\
\multicolumn{2}{l}{2129$+$1210ABHE}
                       &21 29 58.25 .01&   12 10 01.3 0.1& 47633 &    & 21 29 59.27 0.16&$+$12 10 14.1 3.5&$\sim20$&6.3$\pm$0.5& 0.13$\pm$0.51&  ......  &   ......   &?:glbc\\
2139$+$2242 & ......   &21 39 27.00 .30&   22 42 40.0 5.0& 49542 &    & 21 39 26.92 0.13&$+$22 42 39.7 2.0& 1.11&  8.4$\pm$0.5&  1.33$\pm$0.45& 16$\pm$ 5&$-$45$\pm$ 8&     \\
2145$-$0750 & ......   &21 45 50.47 .00&$-$07 50 18.3 0.0& 48979 &10.0& 21 45 50.71 0.41&$-$07 50 15.5 6.7& 4.51&  3.1$\pm$0.5&  0.53$\pm$0.57& 17$\pm$19&   ......   & *    \\
2157$+$4017 &2154$+$40 &21 57 01.81 .01&   40 17 45.8 0.1& 48382 &17.0& 21 57 02.07 0.07&$+$40 17 45.2 0.9& 3.05& 22.2$\pm$0.8&  4.10$\pm$0.49& 18$\pm$ 2&$-$11$\pm$ 3&?:yes    
\\[1mm]
2215$+$1538 & ......   &22 15 39.65 .00&   15 38 34.9 0.1& 49079 &    & 22 15 39.66 0.22&$+$15 38 35.7 2.6& 1.68&  6.0$\pm$0.4&  1.82$\pm$0.50& 30$\pm$ 9&  22$\pm$ 6 &     \\
2219$+$4754 &2217$+$47 &22 19 48.13 .00&   47 54 53.8 0.0& 48382 & 3.0& 22 19 47.32 0.59&$+$47 54 52.1 6.6& 8.33&  3.9$\pm$0.4&  0.39$\pm$0.62& 10$\pm$16&   ......   &     \\
2225$+$6535 &2224$+$65 &22 25 52.36 .02&   65 35 33.8 0.1& 48382 & 2.0& 22 25 52.62 0.97&$+$65 35 38.1 7.0& 4.58&  4.9$\pm$0.5&  2.38$\pm$1.01& 49$\pm$21&  76$\pm$ 6 &     \\
2313$+$4253 &2310$+$42 &23 13 08.57 .01&   42 53 13.0 0.0& 43891 &15.0& 23 13 08.32 0.18&$+$42 53 13.4 1.4& 2.82& 13.2$\pm$1.1&  1.24$\pm$0.53&  9$\pm$ 4&  48$\pm$ 8 &     \\
2330$-$2005 &2327$-$20 &23 30 26.80 .02&$-$20 05 28.6 0.5& 46916 & 3.0& 23 30 26.49 0.37&$-$20 05 36.2 5.8& 8.79&  5.0$\pm$0.5& 2.25$\pm$0.72& 45$\pm$15&$-$27$\pm$ 8 &         
\\[1mm]
2354$+$6155 &2351$+$61 &23 54 04.70 .02&   61 55 46.8 0.1& 48382 & 5.0& 23 54 05.27 0.72&$+$61 55 45.8 5.6& 4.13&  3.0$\pm$0.5&  0.18$\pm$0.53&  6$\pm$18&   ......   &   * \\
\hline
\hline
\vspace{-2mm}\\
\multicolumn{14}{l}{\parbox{230mm}{
Notes~~  ?: more considerations in texts; SNR: confused with supernova remnant; 
no: confused with a strong sources nearby; glbc: confused by host globular cluster;
*: new identifications not presented by Kaplan et al. (1998); mode: PSR B1237+25 is
in abnormal mode as comparing the PA with Fig.7 of Bartel et al. (1982).
}}
\end{tabular}
\end{rotate}
  \end{center}
\end{scriptsize}
\end{table*}
\addtocounter{table}{+1}